\documentclass[sigconf,nonacm]{acmart}
\setcopyright{none}
\usepackage{graphicx}
\usepackage{booktabs}
\usepackage{tabularx}
\usepackage{multirow}

\usepackage{amsmath,amssymb,mathtools}
\usepackage{bm}
\usepackage{siunitx}
\usepackage[linesnumbered,ruled,vlined]{algorithm2e}
\usepackage{bbm}            
\usepackage[normalem]{ulem} 
\usepackage{makecell}
\begin{document}

\title{Urban-STA4CLC: Urban Theory-Informed Spatio-Temporal Attention Model for Predicting Post-Disaster Commercial Land Use Change}

\author{Ziyi Guo}
\affiliation{
  \institution{University of Florida}
  \city{Gainesville}
  \state{FL}
  \country{USA}
}
\email{ziyiguo@ufl.edu}
\orcid{1234-5678-9012}

\author{Yan Wang}
\affiliation{
  \institution{University of Florida}
  \city{Gainesville}
  \state{FL}
  \country{USA}
}
\email{yanw@ufl.edu}

\begin{abstract}
Natural disasters such as hurricanes and wildfires increasingly introduce unusual disturbance on economic activities, which are especially likely to reshape commercial land use pattern given their sensitive to customer visitation. However, current modeling approaches are limited in capturing such complex interplay between human activities and commercial land use change under and following disturbances. Such interactions have been more effectively captured in current resilient urban planning theories. This study designs and calibrates a \uline{U}rban Theory-Informed \uline{S}patio-\uline{T}emporal \uline{A}ttention Model for Predicting Post-Disaster \uline{C}ommercial \uline{L}and Use \uline{C}hange (Urban-STA4CLC) to predict both the yearly decline and expansion of commercial land use at census block level under cumulative impact of disasters on human activities over two years. Guided by urban theories, Urban-STA4CLC integrates both spatial and temporal attention mechanisms with three theory-informed modules. Resilience theory guides a disaster-aware temporal attention module that captures visitation dynamics. Spatial economic theory informs a multi-relational spatial attention module for inter-block representation. Diffusion theory contributes a regularization term that constrains land use transitions. The model performs significantly better than non-theoretical baselines in predicting commercial land use change under the scenario of recurrent hurricanes, with around 19\% improvement in F1 score (0.8763). The effectiveness of the theory-guided modules was further validated through ablation studies. The research demonstrates that embedding urban theory into commercial land use modeling models may substantially enhance the capacity to capture its gains and losses. These advances in commercial land use modeling contribute to land use research that accounts for cumulative impacts of recurrent disasters and shifts in economic activity patterns. 

\end{abstract}


\maketitle

\section{Introduction}
Natural disasters present significant challenges to urban systems by causing damage to physical built environment and affecting human activities\cite{guo_assessing_2025}. The short-term consequences of recurrent or compound shocks can be accumulated and lead to long-term transformations in urban form\cite{safabakhshpachehkenari_japans_2025,nguyen_novel_2024}. This phenomenon is particularly pronounced in commercial land use change in coastal areas. Natural disasters introduce additional challenges, including disrupted physical infrastructure, raised maintenance and insurance cost, and declined customer activities\cite{guo_assessing_2025, guo_benchmarking_2025}. These factors can undermine commercial viability and market value, which potentially leads to temporary or permanent decline in commercial land use if not effectively mitigated. For instance, following Hurricane Ian’s landfall on Florida, U.S. in 2022, coastal commercial districts struggled with prolonged recovery periods, with some areas facing vacancy even after two years \cite{commercial_property_southwest_florida_hurricane_2025}
. 
The decline of commercial areas is often regarded as both a warning sign and a driver of broader cascading deterioration in local economy and vitality\cite{yin_monitoring_2011}. 

With the expectation of recurrent hurricanes in the coastal communities, accurate modeling the impact of natural disasters to land use change is has become essential to meet several critical planning needs. First, disaster preparation planning demands for early detection of systematic vulnerabilities in commercial areas that lie within and beyond traditional hazard maps, which eventually inform efficient resource allocation\cite{guo_assessing_2025}. Second, scenario-based modeling is needed to anticipate continued and unforeseen disaster impacts, informing long-term adaptive land use and economic development planning\cite{hao_empowering_2024}. Third, it is necessary to evaluate the effectiveness of proposed planning strategies aiming at proactive and resilient urban development, such as infrastructure development or zoning adjustments, prior to implementation under disasters impact\cite{linardos_machine_2022}. 

Despite the abundant modelling effort that intend to capture the land use change, the integration of recurrent disaster’s impacts on land use change with a focus on commercial land use has been rare. Changes in commercial land use are unique for two main reasons. First, they exhibit particular sensitivity to fluctuations in customer demand, especially in the context of disasters\cite{guo_assessing_2025, colaco_commercial_2024}. Second, commercial land uses tend to cluster together, forming functional patches\cite{ballantyne_framework_2022}. Their changes are often influenced by spatial spillover effects, including both negative externalities from nearby competition and positive agglomeration economies\cite{hsu_spillover_2025,liu_spatial_2024}. Thus, a specific-designed modelling approach is needed to capture the pathway of disaster impact on commercial land use through economic activity disturbances, built environment disruption among others. Current land use change modeling, which predominantly relies on conventional approaches such as cellular automata (CA) and system dynamic model (SDM), remains limited in its ability to capture changes in built-up land uses or atypical transition pathways\cite{liu_multi-scenario_2021,dede_spatial_2021}. These models also tend to focus on the slow-changing built environment and are insensitive to  temporal changes in visitation within a year, such as weekly or monthly fluctuations, which can also influence land use change. Moreover, while some large-scale models incorporate disasters, they often treat them merely as initial conditions or boundary constraints, rather than fully integrating their dynamic role in shaping land use change\cite{mardalena_model_2024, wang_urban_2023}. 

\begin{figure*}
    \centering
    \includegraphics[width= 0.75\linewidth]{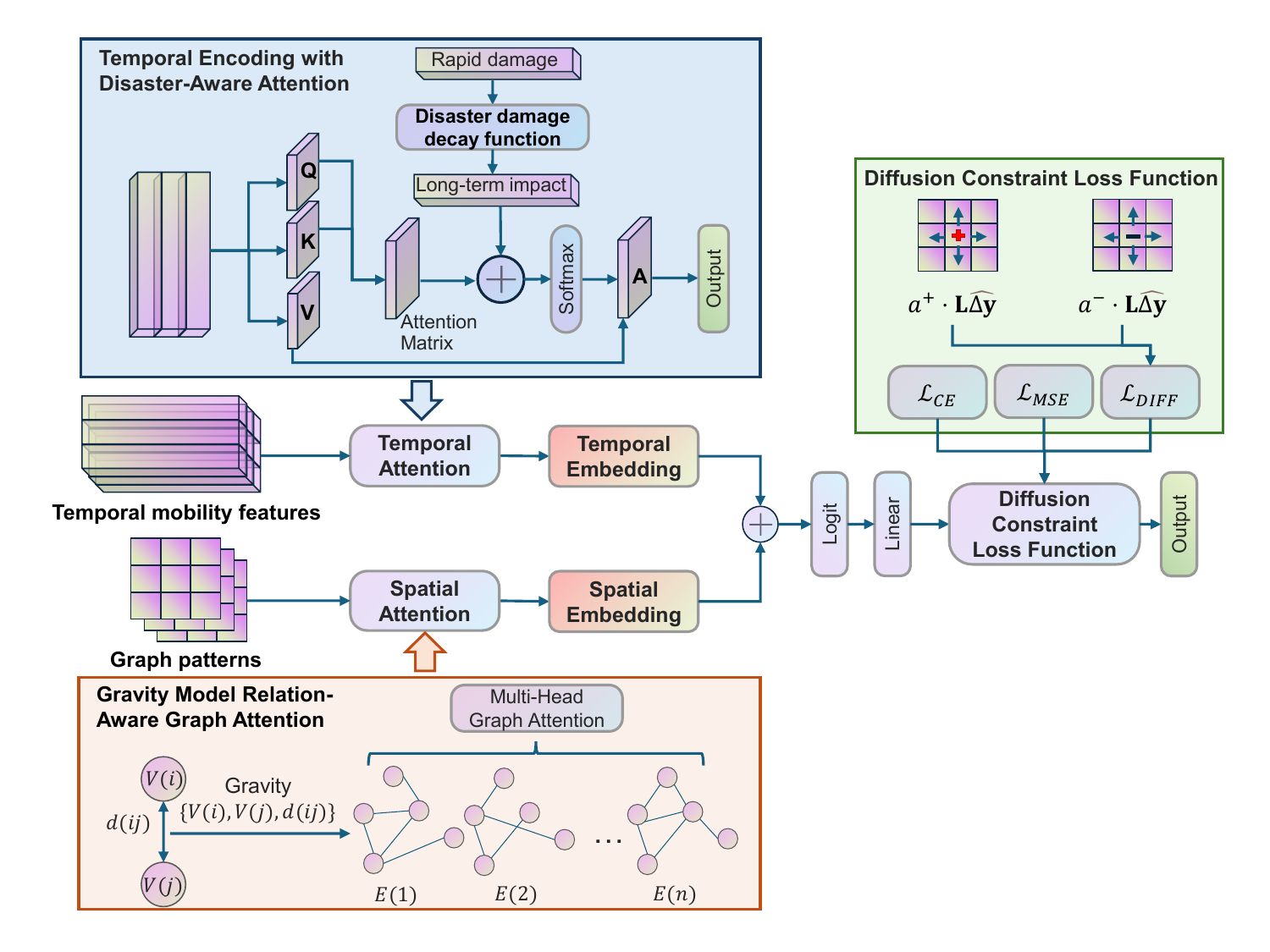}
    \caption{Architecture of Multi-Relational Spatio-Temporal Attentive Network}
    \Description{}
    \label{fig:Figure 1}
\end{figure*}

The growing application of deep learning techniques promise to capture the complex relationship between socio-economic factors, built environments, natural disasters, human activity change and land use patterns across space and time\cite{colaco_commercial_2024,hao_modeling_2023, hou_intracounty_2021, khalid_synergistic_2024}. However, most of the extant efforts are detached from existing planning theories. Given the massive, fluctuating, and heterogeneous nature of urban data, general-purpose deep learning models often face difficulties in extracting meaningful patterns tailored to specific urban challenges that are frequently better understood through theoretical frameworks. In particular, general-purpose models often struggle to capture the complex interplay among disasters, human activity, and land use change due to several key theoretical challenges. The theory of resilience has successfully quantified the impact of disasters using various formulations, such as the exponential decay function\cite{chambers_operationalizing_2019, wang_measuring_2020, wang_urban_2018}. General temporal mechanisms might conflate disaster-induced behavioral shifts with regular fluctuations in visitation. Moreover, spatial economic theories highlight competition for customer visitation among same-sector businesses\cite{doan_attractiveness_2016}. General deep spatial models typically define neighborhoods using physical adjacency alone, overlooking functional linkages between business districts that competition effects\cite{kipf_semi-supervised_2017, velickovic_graph_2018}. Finally, although land use change frequently exhibits spatial diffusion patterns, existing models rarely differentiate between expansion and decline, nor do they enforce spatial coherence through diffusion-based constraints. These limitations highlight the need for a domain theory-informed designs that explicitly integrates economic interdependence, disaster impact accumulation, and directional propagation into a cohesive predictive model\cite{hoffer_theory-inspired_2022, kang_assessing_2023}.

To address these challenges, we propose the Urban Theory-Informed Spatio-Temporal Attention Model for Predicting Post-Disaster Commercial Land Use Change (Urban-STA4CLC), a hybrid model that predicts Census block-level changes in commercial land use. We are among the first to incorporate disaster-induced disruptions to human activities at commercial places within a year as the key impact pathway influencing annual land use change. The model incorporates three key theory-informed innovations.

· First, we established a resilience theory-guided temporal attention mechanism that weight the importance of visitation dynamics, advancing general temporal models by encoding cumulative disaster impacts. 

· Second, we established a multi-relational graph attention mechanism to capture the economic competition effect among commercial places with shared industry classifications, advancing past graph models that only capture spatial adjacency.

· Third, we incorporated a land use diffusion theory-guided constraint into the loss function to model bi-directional spatial propagation, advancing existing by explicitly capturing inter-block externalities.

We calibrate the place-based Urban-STA4CLC model with the data of Cape Coral Metropolitan area in Florida, U.S., where recurrent hurricanes and challenging economic recovery necessitates the modeling of disruption-driven commercial land use change. We trained the model with 260 weeks of data across four time periods with 1,453 census blocks. Each period includes 104 weeks (two years) leading up to the predicted year of land use change. The model performance achieves an F1 of 0.8763, significantly outperforming none-theory informed baselines and ablation variants. The results demonstrate that incorporating disaster decay, competition networks, and directional spatial constraints yields more accurate predictions of post-disaster commercial transformation.

\section{Problem Statement}
We provide a clear definition of our problem in the following section, then give out our solution in next section.
\subsection{Spatio-temporal Settings and Terminology}
We define a spatial graph $\mathcal{G} = (\mathcal{V}, \{ \mathcal{E}_r \}_{r=1}^R)$, where each node $i \in \mathcal{V}$ represents a spatial block and each edge set $\mathcal{E}_r$ encodes a distinct spatial relationship among blocks. Each block $i$ has a fixed spatial location $\bm{l}_i \in \mathbb{R}^2$ and is connected to its neighbors according to each relation type $r$.Each block $i$ is also associated with a vector of node attributes $\bm{z}_i \in \mathbb{R}^d$, which may include long-term characteristics that updated yearly (e.g., business components, built environment configurations, infrastructures and socioeconomic patterns). These attributes determined the probability of land use change and the overall resilience of the place. 

For each block $i$ and each year $s \in { \text{start}, \text{end} }$, we observe a scalar value $y_i^s \in [-1, 1]$, which denotes the percentage of commercial land use in that year. The change in commercial intensity is defined as:
\begin{align}
\Delta y_i = y_i^{\text{end}} - y_i^{\text{start}}.
\end{align}

We then classify each block into one of three categories based on $\Delta y_i$:
\begin{equation}
\label{eq:label}
\hat{y}_i =
\begin{cases}
\text{Increase}, & \text{if } \Delta y_i > \epsilon \\
\text{No change}, & \text{if } |\Delta y_i| \leq \epsilon \\
\text{Decrease}, & \text{if } \Delta y_i < -\epsilon
\end{cases}
\end{equation}

where $\epsilon > 0$ is a small threshold to account for minor fluctuations and noise. The prediction target of our model is to classify each block $i$ into one of the three categories based on the observed features.

Within each block $i$, we observe a collection of point-of-interest (POI) entities indexed by $j$. Each POI $j$ located in block $i$ has an associated weekly visitation time series:
\begin{align}
\bm{v}_{ij} = \{ v_{ij}^1, v_{ij}^2, \dots, v_{ij}^T \} \in \mathbb{R}^T,
\end{align}
where $v_{ij}^t$ is the number of visits to POI $j$ during week $t$. POIs are also typed by business category, which provides semantic context for aggregation and relation inference.

The POI-level visitations are aggregated weekly at the block level to yield a total visitation sequence:
\begin{align}
\bm{v}_i = \left\{ v_i^1, v_i^2, \dots, v_i^T \right\}, \quad \text{where} \quad v_i^t = \sum_{j \in \mathcal{P}_i} v_{ij}^t,
\end{align}
and $\mathcal{P}_i$ denotes the set of POIs located in block $i$. Optionally, we may also track the weekly POI count $p_i^t = |\mathcal{P}_i^t|$ to describe temporal business dynamics.

In addition to visitation data, we also incorporate daily weather and hazard exposure data for each block over the same time window $T$. For each day $t$, we observe a feature vector $\bm{w}_i^t \in \mathbb{R}^{d_w}$ for block $i$, where $d_w$ varies depending on the type of focused hazard (e.g., hurricane, wildfire, extreme heat). The full sequence is denoted as:
\begin{align}
\bm{W}_i = \{ \bm{w}_i^1, \bm{w}_i^2, \dots, \bm{w}_i^T \} \in \mathbb{R}^{T \times d_w}.
\end{align}

These indicators are used to model the immediate and lingering effects of environmental hazards on local visitation patterns and, ultimately, land use change.

\subsection{Land Use Change Prediction Model}

The objective of this study is to learn a predictive model $\mathcal{F}_\theta$ parameterized by $\theta$, which estimates the change in commercial land use $\Delta y_i$ based on:
(1) temporal impacts of rapid and chronic climate and weather extremes; 
(2) temporal patterns of visitation to POIs at the block level;
(3) block-specific attributes that vary over years; and
(4) spatial dependencies encoded in the multi-relational graph of blocks

Formally, the model is defined as:
\begin{align}
\widehat{\Delta y}_i = \mathcal{F}_\theta\left( \bm{v}_i, \bm{W}_i, \bm{z}_i, \{ \bm{v}_j, \bm{W}_j, \bm{z}_j \}_{j \in \mathcal{N}_r(i)}, \bm{A}_r \right),
\end{align}

where $\bm{v}_i \in \mathbb{R}^T$represents weekly visitation sequence for block $i$. $\bm{W}_i \in \mathbb{R}^{T \times d_w}$ is daily weather and hazard feature sequence for block $i$ over $T$ days, including variables relevant to the dominant hazards (e.g., rainfall, wind, fire risk). $\bm{z}_i \in \mathbb{R}^d$ indicates a vector of yearly-updated node attributes for block $i$. $\mathcal{N}_r(i)$ is the set of neighbors of block $i$ under relation type $r$. $\bm{A}_r$refers to the adjacency matrix representing the edge set $\mathcal{E}_r$.
This model is implemented using a hybrid neural architecture, described in Methodology.

\subsection{Modeling Challenges}

We address three domain-specific challenges related to modeling commercial land use change under natural disasters:

\textbf{(1) Temporal emphasis on cumulative disaster impact on economic activities.} Disasters not only cause rapid disturbances in visitation behavior and physical damage to business sites, but also alter economic activities over extended periods\cite{wang_urban_2018}. These changes can accumulate over time, gradually reshaping human activity patterns. Although visitation dynamics may implicitly capture some disaster-induced behavioral shifts, they often fail to separate these effects from regular fluctuations. Moreover, time series visitation signals should not be treated equally when considering their resilience to disasters—post-disaster periods generally carry more weight and are more closely linked to outcomes such as land use change \cite{chambers_operationalizing_2019, wang_measuring_2020}. At the same time, the impacts of disasters on human activities can vary widely across locations within a city, thus requiring localized inputs. Without a theory-informed mechanism to embed and track disaster disturbance effects, models are likely to overlook the temporal signals critical to understanding long-term urban change. 

\textbf{(2) Attention-Based Aggregation of Heterogeneous Graph Structures.} Changes in business activity in commercial locations, along with the related emergence or decline of commercial land use, can propagate beyond adjacent blocks\cite{doan_attractiveness_2016}. Commercial places with similar functions often compete for customer demands, forming latent functional or economic competition networks \cite{kipf_semi-supervised_2017, velickovic_graph_2018}. However, most existing graph-based deep learning models define spatial neighborhoods solely based on geographic adjacency, limiting their ability to capture indirect or non-local relationships. While graph attention mechanisms can weigh the relative importance of neighboring nodes, these weights are typically learned over a single type of edge. In the context of commercial land use, where places are embedded in multiple overlapping relationships (e.g., functional competition of different sectors), a single-relational attention model may fail to capture the varying importance of nodes across different relational dimensions\cite{doan_attractiveness_2016}.

\textbf{(3) Bidirectional Diffusion Dynamics in Land Use Transitions.} Urban development and decline are not random—they tend to diffuse through spatial and functional networks\cite{hoffer_theory-inspired_2022, kang_assessing_2023}. This is particularly true for commercial areas, where agglomeration effects amplify both the growth and retreat of businesses\cite{liu_spatial_2024}. Existing graph-based models rarely consider the diffusion of expansion and decline, nor do they explicitly differentiate between the bidirectional spatial propagation. This limits their ability to model the asymmetrical dynamics of commercial land transformation.

\section{Methodology}
\subsection{Model Design}

To address the aforementioned challenges, we propose the Urban Theory-Informed Spatio-Temporal Attention Model for Predicting Post-Disaster Commercial Land Use Change (Urban-STA4CLC). The overall architecture is illustrated in Figure \ref{fig:Figure 1}. Urban-STA4CLC is composed of three key components. First, the temporal encoder employs an attention mechanism to model sequential dependencies and dynamic trends in the time series inputs. The attention mechanism is engineered to be \textit{resilience theory}-guided to incorporate cumulative disaster impacts, enabling the module to highlight both immediate and delayed behavioral shifts related to disasters. Second, the spatial encoder, built upon a graph attention network (GAT), processes spatial inputs by combining node-level features with multiple layers of relational graphs. A relation-aware attention mechanism is introduced to capture heterogeneous spatial dependencies, including geographic adjacency and sector-specific economic competition networks. This design is informed by \textit{spatial economic theory} to reflect functional linkages among commercial areas. Third, the spatio-temporal fusion and prediction module integrates the learned spatial and temporal embeddings to produce final land use change predictions. To reflect the bi-directional land use propagation, this component extends the loss function with a graph diffusion constraint to reflect incorporating insights from\textit{ land use diffusion theory}.

\subsection{Temporal Encoding with Attention Accounting Disaster Decay Impact }
The temporal input to our model captures weekly dynamics of human activity and environmental stressors over a fixed time horizon of $T$ weeks. For each spatial block $i$, we define a multivariate temporal input matrix:

\[
\bm{X}_i = \left[ \bm{v}_i,\, \bm{p}_i,\, \bm{r}_i,\, \bm{w}_i^{(1)},\, \bm{w}_i^{(2)},\, \bm{w}_i^{(3)} \right] \in \mathbb{R}^{T \times 6},
\]
where:
\(\bm{v}_i \in \mathbb{R}^T\) denotes the weekly POI visitation in block \(i\),
\(\bm{p}_i \in \mathbb{R}^T\) is the weekly count of active POIs in block \(i\),
\(\bm{r}_i \in \mathbb{R}^T\) represents a visitation resilience metric of \(i\),
\(\bm{w}_i^{(k)} \in \mathbb{R}^T\) for \(j = 1,2,3\) are weekly-aggregated weather-related indicators. For the case of hurricane-related stressors, the three weather indicators are precipitation, wind speed, and atmosphere pressure.

The sequence \(\bm{X}_i\) is passed into a temporal encoder composed of a multi-head self-attention mechanism. This encoder computes attention over time to capture latent temporal dependencies and structural shifts in human activity.

In particular, we calculated the visitation resilience metric  \(\bm{r}_i\) based on the theory of resilience for dynamic system, which has been refereed to as the ball-and-basin framework\cite{scheffer_catastrophic_2001, steinmann_resilience_2024}. In this paradigm, the state of a system is conceptualized as a ball moving within a potential landscape. Stable states correspond to the minima (basins) of this potential landscape. Resilience is represented by the steepness (curvature) of the basin around these minima: deeper and steeper basins indicate stronger restoring forces that help the system return to equilibrium after perturbation.

Formally, the dynamics of a one-dimensional system state $y(t)$ can be described as:
\begin{equation}
\frac{dy}{dt} = f(y),
\label{eq:dydt}
\end{equation}
where $f(y)$ is the deterministic component of the system's dynamics. To apply this conceptual framework to our empirical visitation data, we replaced $y(t)$ with $v(t)$.
The corresponding potential function $V(v)$ is obtained by integrating $-f(v)$:
\begin{equation}
V(v) = -\int f(v) \, dv.
\label{eq:potential}
\end{equation}
The curvature of the potential function at its minimum, given by the second derivative $V''(v^*)$, where $y^* = \arg\min V(v)$, serves as a local measure of resilience:
\begin{equation}
\text{Resilience} \propto V''(v^*).
\label{eq:resilience}
\end{equation}
To compute a time series visitation resilience metric, we developed a method to reconstruct the potential landscape over time using rolling windows.  
For each time step $t_i$, we define a temporal window of $T$ centered at $t_i$ and extract the corresponding subseries $\{(t_j, v_j)\}$.  The detailed procedure is described by  Algo. \ref{alg:temporal-resilience}.

\begin{algorithm}[t]
    \caption{Rolling Estimation of Temporal Resilience Metric}
    \label{alg:temporal-resilience}
    \KwIn{Visitation time series $v(t)$, time points $t$, window length $T$, number of bins $B$}
    \KwOut{Temporal resilience index $r(t)$}
    
    Initialize $r(t) \leftarrow$ array of NaNs of same length as $t$\;
    
    \ForEach{$t_i \in t$}{
        Define window $[t_i - T/2, t_i + T/2]$\;
        Extract subseries $(t_j, v_j)$ within window\;
        \If{number of valid points $< 6$}{
            \textbf{continue}\;
        }
        Fit a smoothing spline $s(t)$ to $(t_j, v_j)$\;
        Compute derivative $f(t) = \frac{ds}{dt}$\;
        Evaluate dense grid $\tilde{t}$, get $v_{\text{dense}} = s(\tilde{t})$, $f_{\text{dense}} = f(\tilde{t})$\;
        Bin $y_{\text{dense}}$ into $B$ bins and compute mean $\hat{f}(v)$\;
        Interpolate $\hat{f}(v)$ to estimate continuous $f(v)$\;
        Numerically integrate $-\hat{f}(v)$ to get potential $V(v)$\;
        Compute second derivative $V''(v)$ via finite differences\;
        Identify $y^* = \arg\min V(v)$\;
        Set $r(t_i) \leftarrow V''(v^*)$\;
    }
    \Return $r(t)$\;
\end{algorithm}

\(\bm{r}(t_i)\) are defined as  \emph{temporal resilience metric}. This index captures the evolving capacity of the system to return to a stable visitation state following external or internal perturbations.

In addition, we adopt a temporal exponential decay function to characterize the diminishing impact of a disaster over time~\cite{wang_urban_2018}. Suppose a disaster happens at time $ t_k$, the decayed impact at $t$ after $t_k$ is: 
\begin{equation}
d(t) = d_k e^{-\alpha (t-t_k)},
\label{eq:decay}
\end{equation}
where $\alpha$ is a learnable decay parameter that governs the rate of disturbance decline, and $d_k$ is the initial magnitude of disaster-induced disturbance, assumed to peak at $t = t_k$. As time progresses, $d(t)$ approaches zero. In the case of recurrent disasters, the cumulative effects are added at each $t$, forming a final disturbance sequence \(\bm{D}_i \in \mathbb{R}^T\).
Figure 2 illustrates the conversion of timestamped disaster events into a time-series representation of cumulative disaster impact. With a larger                     , the influence of each disaster decays more rapidly (Fig.2.a); whereas small    allows the effects of previous disasters to persist and accumulate when subsequent disasters occur before earlier impacts have fully decayed (Fig.2.b).

\begin{figure}
    \centering
    \includegraphics[width=1\linewidth]{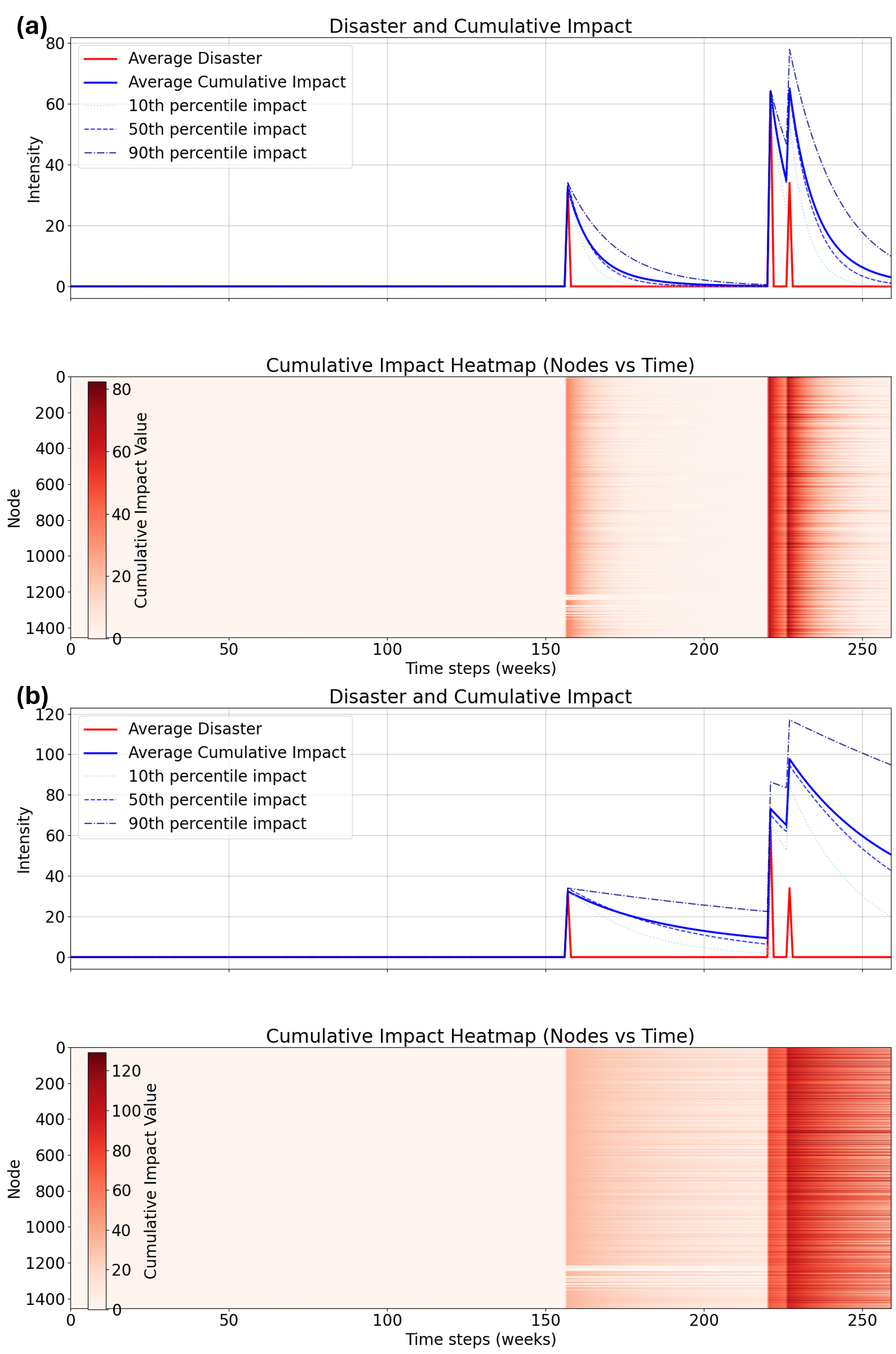}
    \caption{Example of cumulative decayed impact of disasters under the different decay parameter (a) decay parameter of 0.5 indicates a rapid impact; (b) decay parameter of 0.05 indicate a prolonged impact}
    \Description{}
    \label{fig:enter-label}
\end{figure}

To inform the attention mechanism with disaster impacts, we incorporate \(\bm{D}_i \)  as a residual that is injected into the attention logits as an additive bias. For each block \(i\), temporal self-attention is calculated as follows:
\begin{equation}
\text{Attn}(Q, K) = \text{Softmax}\left( \frac{QK^\top}{\sqrt{d}} + \bm{D}_i \right)V,
\end{equation}
where the bias term \(\bm{D}_i\) is broadcasted to align with the attention score tensor. This allows the model to prioritize time steps with stronger post-disaster effects and to learn location-specific sensitivity to disruption. The result of temporal encoder \(\bm{h}_i^{\text{temp}}\) captures both visitation dynamics and weather and disaster disturbances. 

\subsection{Spatial Encoding with Relation-Aware Graph Attention}

To model the spatial dependencies that influence land use change, we construct a multi-relational graph \(\mathcal{G} = (\mathcal{V}, \{ \mathcal{E}_r \}_{r=1}^R)\), where each node \(i \in \mathcal{V}\) represents a spatial block. Two types of edge relations are defined to capture spatial and economical relationships among business places.

The first relation type is based on spatial proximity. For each block \(i\), we identify its 10 geographically nearest neighbors to construct an undirected edge set \(\mathcal{E}_{\text{adj}}\), referred to as the \textit{adjacency baseline}. Instead of assigning uniform edge weights, we define the weight of each edge as the inverse of the Euclidean distance between the centroids of the two connected blocks, so that closer neighbors exert stronger influence.

The second relation type captures inter-block economic competition between POIs in the same industry sector. Let \(S_k\) denote one of the 67 three-digit NAICS industry sectors, among which 53 of them have more than 20 blocks. For each sector \(k\), we define a sector-specific edge set \(\mathcal{E}_k\), which connects blocks that both contain POIs belonging to \(S_k\). The weight of a competitive linkage between blocks \(i\) and \(j\) under sector \(k\) is defined using a gravity-inspired formulation:
\begin{equation}
C_{ij}^{(k)} = \frac{V_i^{(k)} \cdot V_j^{(k)}}{d_{ij}^2},
\end{equation}
where \(V_i^{(k)}\) is the total number of visits to POIs in sector \(k\) located in block \(i\), and \(d_{ij}\) is the geographic distance between blocks \(i\) and \(j\). This formulation follows classical principles of spatial interaction theory~\cite{liang_calibrating_2020}. 

These two types of relations are used to construct a multi-relational graph structure with \(R = 54\) relation types. We encode spatial structure using a relation-aware graph attention network. For each relation \(r\), a dedicated edge-aware attention module computes node embeddings based on node features \(\bm{h}_i\) and edge features \(\bm{e}_{ij}^{(r)}\). The output of all relations is aggregated using a relation-level soft-attention mechanism. The resulting block-level spatial embedding \(\bm{h}_i^{\text{spatial}}\) captures both spatial adjacency and industrial competitive dynamics:
\begin{equation}
\bm{h}_i^{\text{spatial}} = \sum_{r=1}^{R} \alpha_r^{(i)} \cdot \text{GAT}^{(r)}(\bm{h}_i, \mathcal{E}_r, \bm{e}_{ij}^{(r)}),
\end{equation}
where \(\alpha_r^{(i)}\) denotes the learned relation attention weight for relation \(r\) at node \(i\). This flexible structure enables the model to adaptively balance different types of spatial signals relevant to changes in commercial land use.

\subsection{Training Objective with Diffusion Constraint}

After obtaining the temporal embedding \(\bm{h}_i^{\text{temp}}\) from the disaster-aware attention encoder and the spatial embedding \(\bm{h}_i^{\text{spatial}}\) from the relation-aware GAT, we concatenate the two vectors and project them through a fully connected layer:
\begin{equation}
\bm{h}_i^{\text{fused}} = ReLU \left( \bm{W}_f \left[ \bm{h}_i^{\text{temp}} \, \Vert \, \bm{h}_i^{\text{spatial}} \right] + \bm{b}_f \right),
\end{equation}
where \(\bm{W}_f\) and \(\bm{b}_f\) are learnable parameters, \(\Vert\) denotes vector concatenation, and \(ReLU\) is a non-linear activation. 

The fused representation is passed through a regression head with a \(\tanh\) activation to produce the predicted land use change:
\begin{equation}
\widehat{\Delta y}_i = \tanh\left( \bm{W}_{\text{out}} \bm{h}_i^{\text{fused}} + b_{\text{out}} \right),
\end{equation}
where the output is constrained to lie within \([-1, 1]\), consistent with the domain of the target variable: the change in percentage of commercial land use.

We incorporate domain knowledge on the spatial propagation of commercial land use, where agglomeration effects cause growth areas to attract further development and decline areas to lose more activity. These reinforcing dynamics align with diffusion-like patterns, justifying the use of a spatial diffusion constraint.  Diffusion models (e.g., reaction–diffusion and heat diffusion dynamics) have been widely adopted as proxies to describe the process of urban boundary expansion and land use transitions ~\cite{ohtake_pattern_2025, liu_quantifying_2023}. Inspired by this theory, we introduce a simplified bidirectional diffusion constraint that captures both the expansion and contraction of commercial land use over space.

The spatial diffusion process is implemented using a discrete graph Laplacian \(\bm{L}\), derived from the spatial adjacency graph. We define the residual diffusion dynamics at each node \(i\) as:

\begin{equation}
\text{Res}_i = 
\begin{cases}
\widehat{\Delta y}_i - a^{+} \cdot (\bm{L} \widehat{\Delta \bm{y}})_i, & \text{if } \widehat{\Delta y}_i > 0 \\
\widehat{\Delta y}_i - a^{-} \cdot (\bm{L} \widehat{\Delta \bm{y}})_i, & \text{if } \widehat{\Delta y}_i < 0
\end{cases}
\end{equation}

where \(a^{+}\) and \(a^{-}\) are learnable diffusion coefficients representing the rate of commercial expansion and decline, respectively. This formulation enables the model to impose directional spatial consistency based on the observed type of land use transition.

where \(a^{+}\) and \(a^{-}\) are learnable diffusion coefficients for increasing and decreasing land use transitions, respectively. The total diffusion loss is:
\begin{equation}
\mathcal{L}_{\text{Diff}} = \lambda_{\text{Diff}} \left( \frac{1}{N_{+}} \sum_{i:\widehat{\Delta y}_i > 0} \text{Res}_i^2 + \frac{1}{N_{-}} \sum_{i:\widehat{\Delta y}_i < 0} \text{Res}_i^2 \right),
\end{equation}
where \(N_{+}\) and \(N_{-}\) are the number of positively and negatively changing blocks, respectively, and \(\lambda_{\text{PDE}}\) is a hyperparameter controlling the strength of the diffusion constraint.

The final training objective combines three components: a categorical loss for predicting the direction of change, a regression loss for estimating the magnitude of change, and a physics-informed loss for enforcing spatial diffusion constraints.

We first define the categorical loss \(\mathcal{L}_{\text{cls}}\) using cross-entropy over three classes—\textit{increase}, \textit{no change}, and \textit{decrease}—based on the sign and magnitude of the true change \(\Delta y_i\). The regression loss \(\mathcal{L}_{\text{reg}}\) is defined as the mean squared error (MSE) between the predicted and true values:
\begin{equation}
\mathcal{L}_{\text{reg}} = \frac{1}{N} \sum_{i=1}^{N} \left( \widehat{\Delta y}_i - \Delta y_i \right)^2.
\end{equation}

In addition, we incorporate a physics-informed loss \(\mathcal{L}_{\text{Diff}}\) to reflect spatial diffusion dynamics based on domain knowledge.

The total loss function is:
\begin{equation}
\mathcal{L}_{\text{total}} = \lambda_1 \cdot \mathcal{L}_{\text{cls}} + \lambda_2 \cdot \mathcal{L}_{\text{reg}} + \lambda_3 \cdot \mathcal{L}_{\text{Diff}},
\end{equation}
where \(\lambda_1\), \(\lambda_2\), and \(\lambda_3\) are weighting coefficients that balance the contribution of each component.

The model is trained end-to-end using stochastic gradient descent with the Adam optimizer. All parameters, including those in the predictive network and the physical constraints (e.g., \(a^+, a^-\)), are jointly optimized.

\section{Results}
\subsection{Case and Data Description}

\begin{figure*}
    \centering
    \includegraphics[width=0.75\linewidth]{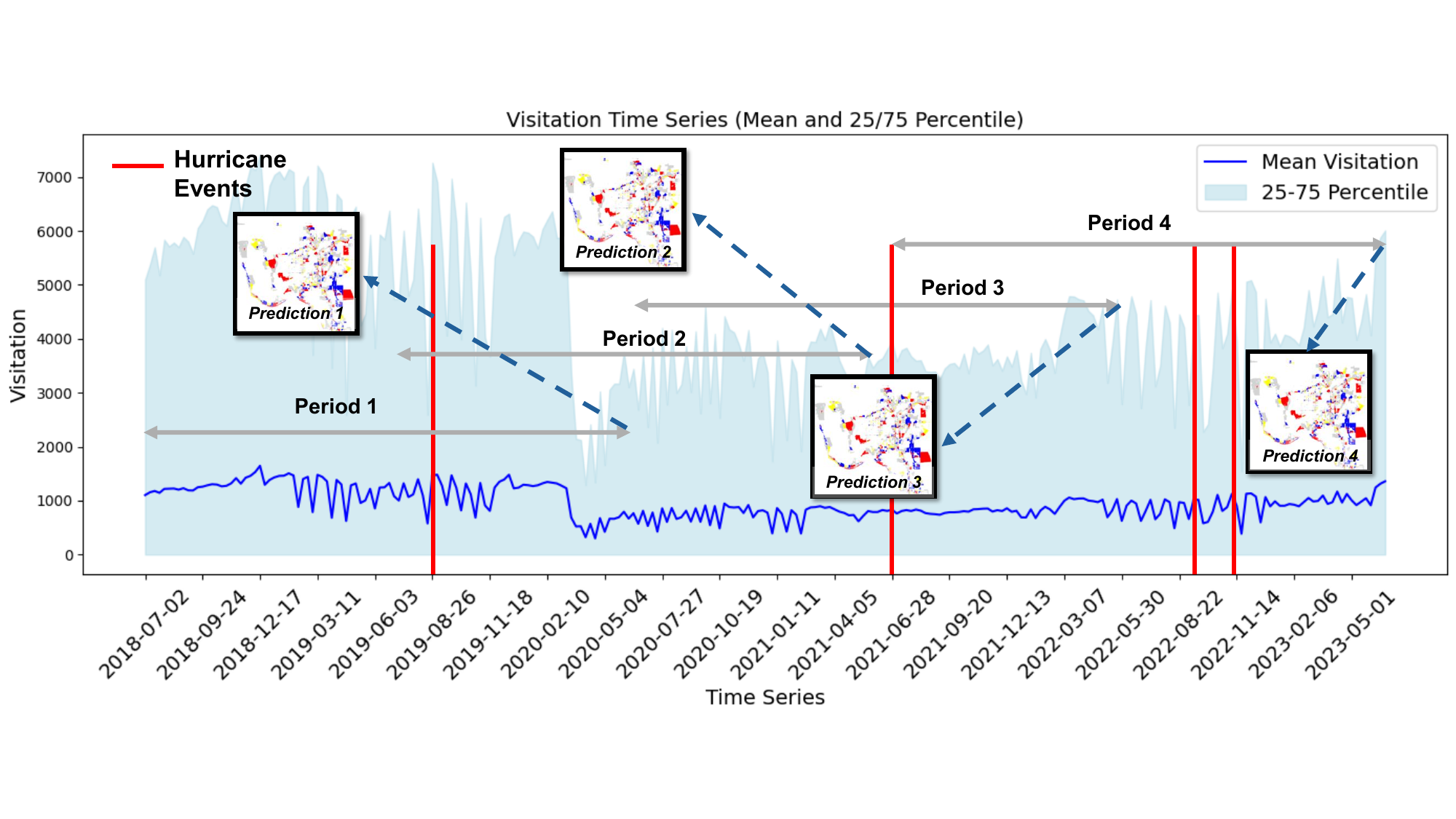}
    \caption{Illustration of training data and division of periods}
    \Description{}
    \label{fig:3}
\end{figure*}

We calibrated the Urban-STA4CLC model for the Cape Coral metropolitan area, located along the Gulf Coast of the state of Florida, U.S., spanning Lee County and Charlotte County, Florida. Weekly record at POIs from July 2018 to June 2023, across 260 weeks in total have been used for model calibration. The training dataset consists of four time periods, each spanning two years—from July of the start year to June of the end year. For each period, we predict land use change in the end year based on visitation dynamics over the preceding two years \ref{fig:3}. 
We model commercial land use change at the U.S. Census Block level, leveraging its compatibility with both data availability and real-world urban structures. Our dataset integrates multi-source inputs. Although fine-grained, pixel-based land use models are widely applied, we focus on block-level commercial land use change for several reasons. First, the development of commercial areas tends to align with real-world urban structures and boundaries. Census blocks, which are defined by physical features such as roads and rivers, better reflect actual land use patterns than regular grid-based pixels. Second, the contextual features we employ are generally unavailable or unreliable at fine pixel resolutions (e.g., 10–30 meters), but are systematically collected at the census block level\cite{bureau_american_2023}. Third, while detailed zoning and development control are often conducted at the parcel level, broader planning efforts—such as infrastructure development, disaster mitigation, and economic development planning—typically operate at the block or higher geographic scales. Lastly, considering the agglomerative nature of businesses and the patch-based functional structure of commercial places, pixel-level modeling often suffers from excessive fragmentation. Modeling at the census block level provides a smoother spatial representation that better captures macro-scale commercial dynamics.

Among more than 4,000 blocks in the study area, 1,453 contained nonzero commercial land use at either the beginning or end of the study period. Since a large proportion of blocks exhibited either no change or only minimal change in commercial land use, we applied resampling techniques to balance the distribution of samples across different classes for the training dataset. We used an absolute percentage land use change threshold of 1×10\textsuperscript{-5 }to distinguish between blocks with changed and unchanged commercial land use.

\subsection{Feature Description}

\textit{\textbf{Temporal Features}} The core temporal input is the sequence of visitation changes observed at each block over time. We obtain POI location records and block-level visitation data from Advan\cite{dewey_advan_2025}
. 
Each POI includes a North American Industry Classification System (NAICS) code, which designates its business type. Based on the 3-digit NAICS level, we group POIs into 67 distinct industrial sectors. These groupings form the basis for constructing sector-specific inter-block competition networks. In addition to the original visitation data, we incorporate weekly counts of active businesses and a calculated resilience index as supplementary predictors (see Method). Lastly, we include a separate temporal sequence documenting of continued weather conditions\cite{customweather_daily_2025} 
and all disasters that have occurred in the study area\cite{fema_disaster_2025}
. The significance of each event is calculated at the neighborhood level and is further processed into a cumulative disaster impact variable through feature engineering (see Method). 
\textit{\textbf{Spatial Features}}. The block-level characteristics incorporate a rich set of 38 features encompassing built environment metrics, land use composition, business structure, and neighborhood socioeconomic characteristics \ref{tab:spatial_features}. All variables are standardized and input to the model as static node attributes in the spatial graph. These features are selected based on prior evidence linking them to both human activity patterns and land use dynamics.

\subsection{Implementation}

All models are implemented in PyTorch with CUDA 12.0 and trained on a single GPU with 5~GB memory. Given the moderate size and sparsity of the graph, we opt for full-batch training rather than mini-batch sampling. To ensure fair comparison across all model variants, we fix the random seed so that each training run starts from the same initialization.

The dataset was trained and validated using a five-fold cross-validation approach, with each fold randomly partitioned into training (80\%) and validation (20\%) sets. All learnable layers were configured with a hidden dimension of 64. The learning rate was fixed at 0.005 throughout the training process. The weight for the diffusion loss term, $\lambda_{\mathrm{diff}}$, is set to 0.05 based on grid search over a validation set.

\subsection{Model Evaluation}
\begin{figure}
    \centering
    \includegraphics[width=1\linewidth]{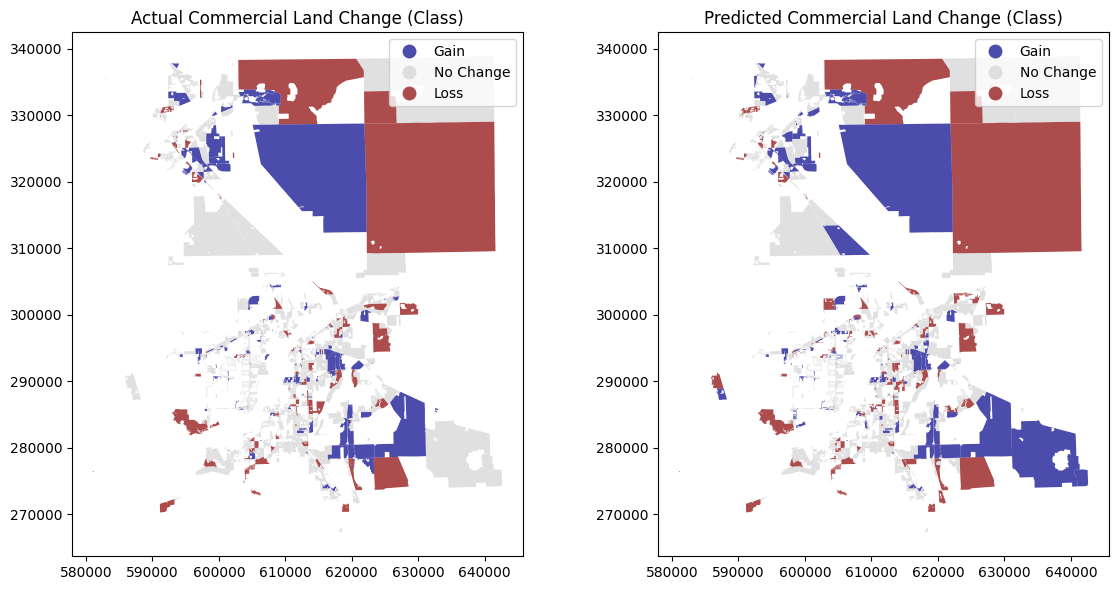}
    \caption{Predicted and actual land use change in year 2021-2023}
    \Description{}
    \label{fig:Figure 3 map}
\end{figure}

Figure 3\ref{fig:Figure 3 map} illustrates the spatial comparison between the predicted and observed commercial land use changes across the study area. We evaluate the performance of the proposed model using four standard metrics: best loss, F1 score, precision, and recall, which together provide a comprehensive assessment of predictive accuracy and robustness. Given the class imbalance in the data, we selected F1 score as the primary evaluation metric. To establish a comparative benchmark, we consider two baseline models: Spatio-Temporal Graph Convolutional Networks (STGCN), which incorporates both spatial and temporal inputs; and Spatio-Temporal Attentive Network (STAN), the base attention-driven spatiotemporal model prior to the integration of theory-guided modules. We focus exclusively on machine learning approaches for comparison, as conventional land use change models are not designed to accommodate the high-resolution spatiotemporal inputs or dynamic disaster impact data used in our study. 
Model performance comparisons demonstrate the value of incorporating domain-informed components. Among the baselines, STGCN performs the weakest (F1: 0.1074) and is included for reference. The STAN model, which uses the same core architecture as Urban-STA4CLC but excludes the disaster-aware attention module, diffusion constraint, and competition network, performs moderately (F1: 0.7375; precision: 0.7402; recall: 0.7354). While it maintains a balanced evaluation metrics, the performance highlights the limitations of relying solely on generic spatiotemporal features without domain-informed mechanisms. 

Urban-STA4CLC demonstrates strong and consistent performance across cross-validation folds. The model achieves an average training F1 score of (95.9\%) and accuracy of (95.8\%), indicating effective convergence and minimal overfitting. On validation sets, the model maintains robust generalization, with an average F1 score of (87.7\%), precision of (88.6\%), recall of (87.6\%), and accuracy of (87.5\%). Confusion matrices across folds show balanced predictions across classes, with limited variance in performance, further supporting the model’s reliability in multi-class post-disaster land use change prediction.

\begin{table}[t]
\centering
\caption{Evaluation of performance for Urban-STA4CLC and baseline models}
\label{tab:performance_evaluation}
\begin{tabularx}{\linewidth}{>{\hsize=1\hsize\linewidth=\hsize}Xcccc}
\toprule
\textbf{Model Configuration} & \textbf{Training Loss} & \textbf{F1} & \textbf{Precision} & \textbf{Recall} \\
\midrule
STGCN             & 1.1370 & 0.1074 & 0.0592 & 0.6667 \\
STAN (baseline)   & 0.0164 & 0.7375 & 0.7402 & 0.7354 \\
Urban-STA4CLC     & 0.1092 & 0.8763 & 0.8868 & 0.8751 \\
\bottomrule
\end{tabularx}
\end{table}

\subsection{Ablation Study and Component Effectiveness}

To evaluate the contribution of each theory-guided modules, we conduct an ablation study by selectively disabling each or pairs of the novel designs. Eight variants of models are included in the analysis, which are designed to isolate the contribution of disaster-aware attention module, multi-relational competition networks, and the diffusion constraint on loss function. \ref{tab:ablation}summarizes the eight model configurations and corresponding performance. 

The full model (M8), which integrates all three components, achieved the highest validation performance (F1: 0.8763), representing a 18.8\% improvement over the baseline model (M1, F1: 0.7375). This result highlights the effectiveness of combining spatial, temporal, and physics-informed design elements in the classification task.

To examine the impact of each individual module, we first considered models that include only one theory-guided component. Adding the diffusion constraint alone (M2) led to a 10.1\% improvement (F1: 0.8121). Incorporating only the multi-relational competition network (M3) resulted in a 5.7\% gain (F1: 0.7798), while using only the disaster-aware attention module (M4) produced a 5.0\% increase (F1: 0.7740). We further analyzed the performance drop when removing each component from the full model. Excluding the competition network (M6) caused a -9.2\% decrease relative to M8. Removing the disaster-aware attention module (M7) led to an -11.8\% drop, while omitting the diffusion constraint (M4) resulted in an -11.7\% decline.

In summary, the ablation study confirms that each theory-informed module contributes meaningfully to classification accuracy. The full model, which integrates all three, consistently outperforms partial variants and establishes the strongest predictive capability. Among the three theory-informed module, the diffusion constraint provides the greatest standalone benefit. 

\begin{table*}[t]
\centering
\caption{Model performance comparison with respect to baseline STAN configuration}
\label{tab:ablation}
\begin{tabular}{lccccccc}
\toprule
\textbf{Model} & \textbf{Train Loss} & \textbf{Train F1} & \textbf{Val F1} & \textbf{Improvement} & \makecell{\textbf{Disaster} \\ \textbf{Impact}} & \makecell{\textbf{Diffusion} \\ \textbf{Constraint}} & \makecell{\textbf{Multi-relation} \\ \textbf{Network}} \\
\midrule
M1 (Baseline)     & 0.0164 & 0.9677 & 0.7375 & Baseline & No  & No  & No  \\
M2                & 0.0613 & 0.9667 & 0.8121 & 1.1011   & No  & Yes & No  \\
M3                & 0.0060 & 0.9966 & 0.7798 & 1.0572   & No  & No  & Yes \\
M4                & 0.0417 & 0.9612 & 0.7740 & 1.0494   & Yes & No  & No  \\
M5                & 0.0051 & 0.9966 & 0.8174 & 1.1082   & No  & Yes & Yes \\
M6                & 0.0029 & 1.0000 & 0.7955 & 1.0785   & Yes & Yes & No  \\
M7                & 0.0042 & 0.9983 & 0.7733 & 1.0485   & Yes & No  & Yes \\
M8 (Full Model)   & 0.1092 & 0.9613 & 0.8763 & 1.1882   & Yes & Yes & Yes \\
\bottomrule
\end{tabular}
\end{table*}

\section{Conclusion}
In this paper, we present an initial implementation of a theory-informed model, Urban-STA4CLC, tailored to predict atypical changes in commercial land use driven by the cumulative impacts of environmental disasters on economic activities. Through innovative design choices in feature engineering, model architecture, and loss function configuration—guided by theories of disaster resilience, spatial economics, and land use transition—Urban-STA4CLC achieves high predictive accuracy in modeling commercial land use change. Applied to the Cape Coral metropolitan area, the evaluation and ablation analysis confirms the advancement of theory-informed model design and contribution of each innovative module. This work represents a preliminary attempt to integrate well-established urban theories into AI-based urban modeling. The results demonstrate that theory-informed design can enhance model performance. Future work will further develop the model to capture more complex interactions among land use types and population dynamics, with the goal of enabling medium- to long-term forecasting of urban change.
The developed model can be applied to several important planning tasks with additional input and calibration. It can help identify commercial areas at risk of vacancy after one or a sequence of disaster, with the basic understanding of the context of the place and input of disaster impact. In addition to near-term forecasting, Urban-STA4CLC is well-suited for scenario-based planning that facilitates anticipatory decision-making. Two key scenario types can be explored. The first involves recurrent disaster scenarios, which are critical for communities expected to experience frequent disasters over time. Urban-STA4CLC enables simulation of atypical, medium- to long-term commercial land use changes under recurrent shocks. Such scenario modeling can support decisions on long-term economic development, disaster mitigation, and adaptive land use planning to reduce systemic disruptions. The second use case involves evaluating the effectiveness of resilience-oriented planning strategies. By adjusting spatial context variables, such as infrastructure layout, land use patterns, or policy interventions, the model can project how such changes might influence the trajectory of commercial land use in response to natural disasters and shifting human behavior. This capacity allows planners to test and refine land use and infrastructure policies before implementation, making Urban-STA4CLC a powerful decision support tool for building more resilient urban systems.

\begin{table*}[t]
\centering
\caption{Spatial features describing contexts of census blocks}
\label{tab:spatial_features}
\begin{tabular}{lp{10.5cm}}
\toprule
\textbf{Attribute Name} & \textbf{Description} \\
\midrule
\multicolumn{2}{l}{\textit{Socio-economic characteristics}} \\
total\_population & Total population. \\
median\_income & Median household income. \\
prop\_non\_white & Proportion of residents identifying as non-White. \\
prop\_under\_18 & Proportion of the population under the age of 18. \\
prop\_65\_plus & Proportion of the population aged 65 and older. \\
prop\_high\_edu & Proportion of adults with a bachelor’s degree or higher. \\
\midrule
\multicolumn{2}{l}{\textit{General built environment}} \\
avg\_LNDVAL & The value of the land on the parcel. \\
avg\_JV & Total Just Value (land just value plus building value plus special feature value) of the property. \\
Unit\_JV & Unit just value. \\
avg\_ACTYRBLT & Average year of built. \\
sum\_TOTLVGAREA & Total Living Area – Residential OR Adjusted Usable – Non-Residential. \\
sum\_NOBULDNG & Number of buildings. \\
shannon\_diversity & Diversity of land use. \\
\midrule
\multicolumn{2}{l}{\textit{Commercial built environment}} \\
com\_LNDVAL & The value of the land on the parcel of commercial land use. \\
com\_JV & Total Just Value of the commercial property. \\
com\_year & Average year of built for commercial buildings. \\
com\_building & Number of commercial buildings. \\
com\_area & Commercial area. \\
\midrule
\multicolumn{2}{l}{\textit{Land use component}} \\
AGRICULTURAL & Land primarily used for farming, cultivation of crops, raising livestock, or related agricultural activities. \\
CENTRALLY ASSESSED & Properties assessed by a central authority (often state-level), typically including utilities, railroads, etc. \\
INDUSTRIAL & Land used for manufacturing, production, assembly, warehousing, and heavy industry. \\
INSTITUTIONAL & Land used by non-profit organizations, government agencies, or major institutions like schools, hospitals, etc. \\
MINING & Land used for the extraction of minerals, stone, or other geological resources. \\
PARCELS WITH NO VALUES & Land parcels for which property value data is missing or not applicable in the dataset. \\
PUBLIC/SEMI-PUBLIC & Land owned or used by government bodies or organizations serving the public good (parks, libraries, etc.). \\
RECREATION & Land designated or used for recreational purposes, including parks, sports fields, and leisure areas. \\
RESIDENTIAL & Land used primarily for housing, including single-family homes and multi-unit residential buildings. \\
RETAIL/OFFICE & Land used for commercial businesses, such as retail stores, restaurants, and office buildings. \\
ROW & Right-of-Way; land used for transportation corridors (roads, railways) or utilities (pipelines, power lines). \\
VACANT NONRESIDENTIAL & Undeveloped land designated or zoned for non-residential uses (commercial, industrial, etc.). \\
VACANT RESIDENTIAL & Undeveloped land designated or zoned for residential use. \\
WATER & Areas primarily consisting of bodies of water (lakes, rivers, significant ponds) within the boundaries. \\
OTHER & A residual category for land uses that do not fit into any of the specific classifications provided. \\
\bottomrule
\end{tabular}
\vspace{1mm}

\begin{flushleft}
\footnotesize
\vspace{-4pt}
\textit{Note.} Socio-economic features are derived from the American Community Survey 5-Year Data~\cite{bureau_american_2023}. General and commercial built environment features are based on the Florida Statewide Parcels dataset~\cite{commercial_property_southwest_florida_hurricane_2025}. Land use classifications follow GEOPLAN Generalized Land Use definitions~\cite{the_florida_geographic_data_library_fgdl_geospatial_2023}.
\end{flushleft}
\end{table*}

\bibliographystyle{ACM-Reference-Format}
\bibliography{references}

\end{document}